# NANOINDENTATION-INDUCED PHASE TRANSFORMATION IN SILICON

R. Rao,  J. E. Bradby  and  J.S. Williams

Department of Electronic Materials Engineering, RSPhysSE, Australian National University,
ACT 0200, Australia

## ABSTRACT

Nanoindentation-induced phase transformation in silicon has been studied. A series of nanoindentations were made with the sharp diamond Berkovich tip. During nanoindentations, maximum load ranged from 2000 µN to 5000 µN, with a 1000 µN/sec loading rate. Slow unloading rate at 100µN/sec was chosen to favor the formation of the crystalline end phases, high pressure phase (Si-III and Si-XII). Fast unloading rate at 1000µN/sec was used to obtain amorphous phase. The phase transformation was examined by Raman spectroscopy and plan-view transmission electron microscopy (TEM). HPP have been identified even if no "pop-in" and "pop-out" observed in load-depth characteristics curves. HPP appeared in c-Si when the maximum load up to 3000 µN. TEM images have been revealed that the optimization HPP transformation in c-Si at the nanoscale occurred when the maximum load applied at 5000 µN.

## 1. INTRODUCTION

It has been known that the cubic diamond structure of silicon (Si-I) can transform to a metallic phase (Si-II) under indentation load. Upon pressure release Si-II undergoes further transformation. The structure of the final phase can be either a-Si or poly-Si (Si-III/Si-XII). Fast unload rates result in the amorphous phase and slower unload rates in Si-III/Si-XII. Many works have reported the phase transformation in silicon by microindentation [1-3]. Recently, although a few works have examined the phase transformation by nanoindentation [6-7], the mechanisms of the phase transformations and the corresponding microstructural changes in silicon by nanoindentation have never been explored in detail. Obviously, it is significant to study characterization of such phase transformation in Si at the nanoscale.

In this work, we studied characterization of such phase transformation in crystalline silicon (c-Si) at the nanoscale. We focused on the behavior of c-Si under sharp Berkovich indentation during loading and unloading. We analyzed the microstructure changes by Raman spectroscopy and examined the deformation regions using plan-view TEM.

## 2. EXPERIMENT

In the experiments, p-type c-Si wafers with (100) crystallographic orientation and resistivity of 8-12 Ωcm were used. A series of nanoindentations were performed with the sharp diamond Berkovich tip using TriboIndenter (Hysitron Inc., USA). During nanoindentations, maximum load ranged from 2000 µN to 5000 µN, with a 1000 µN/sec loading rate. Slow unloading rate at 100µN/sec was chosen to favor the formation of the crystalline end phases, Si-III and Si-XII. Fast unloading rate at 1000µN/sec was used to obtain amorphous phase.

After indentations, Raman spectra were recorded with Renishaw 2000 Raman Imaging Microscope using the 647.1nm excitation lines of a helium-neon laser focused to a spot of ~ 1 µm radius. According to the atomic force microscopy (AFM) image, the width of the nanoindents is about 500 nm. Therefore, nanoindents closely spaced with 0.5 µm in an effort to obtain enough Raman signal to examine the phase changed. Furthermore, the deformed regions were investigated by plan-view TEM.

## 3. RESULTS AND DISCUSSION

Figure 1 shows a load versus penetration plot for a range of maximum loads. It is well known that c-Si exhibits a discontinuity on loading and unloading during micro-indentation. The discontinuity on loading is called "pop-in" and the discontinuity on unloading is called "pop-out". Such discontinuities during micro-indentation have been reported to occur for c-Si as a result of pressure-induced phase transformations [2]. However, under the nanoindentation conditions shown in Fig.1, no discontinuities on either the loading or unloading sections of the curve are observed and it is interesting to examine if phase transformation have occurred in these cases.

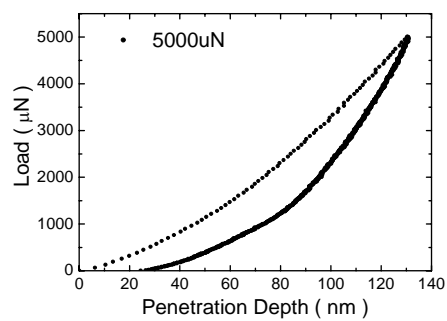

Fig.1 Load-depth curves of nanoindentation in c-Si

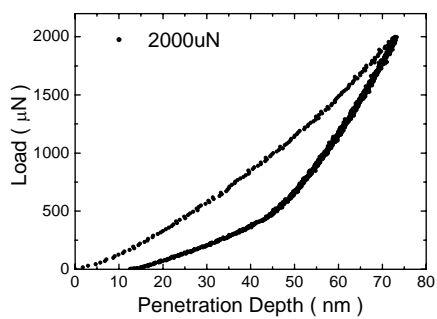

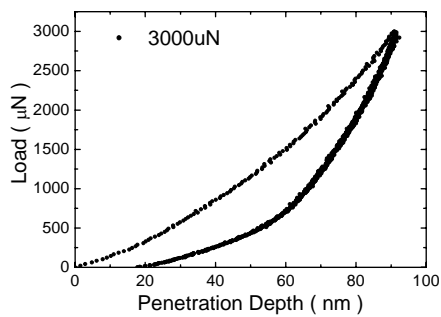

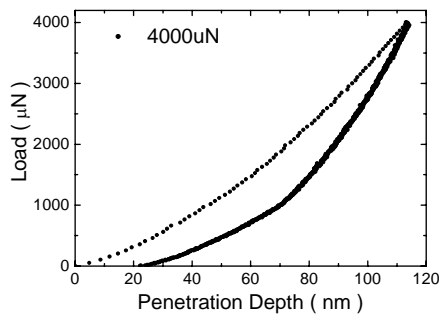

We use Raman spectroscopy (shown in Fig. 2) to examine if phase transformation occurred during these nanoindentations. The multiple narrow bands at 161, 350 382, 394 and 430 $cm^{-1}$ suggest the presence of the SI-III and Si-XII phases. The bands at 350 and 394 $cm^{-1}$ have been assigned to Si-XII, and the bands at 166, 382, 430 $cm^{-1}$, to Si-III. We can see that the Si-III/Si-XII phases start to appear when the maximum load applied reaches about 3000 μN. The Si-III/Si-XII peaks become sharper and stronger with increasing maximum load. Therefore, it is clear that phase transformation can occur in c-Si during nanoindentation even if no discontinuities on either the loading or unloading curve are observed.

Even if we observed HPP peaks appeared at maximum load applied up to 3000 μN in Raman spectra, TEM images (Fig. 4) taken from this deformation region show the nanoindent is amorphous state. However, the SAD pattern from the indent made by 5000 μN shown in Fig. 4(b) is in agreement with the Raman result. Compared the two bright field images shown in Fig. 3(a) and Fig. 4(a), it is obviously that the plump shape of the indents can be obtained at 5000 μN load value. According to the plan-view TEM results, we think that the optimization high pressure phase transformation in c-Si at the nanoscale occurred when the maximum load applied at 5000 μN.

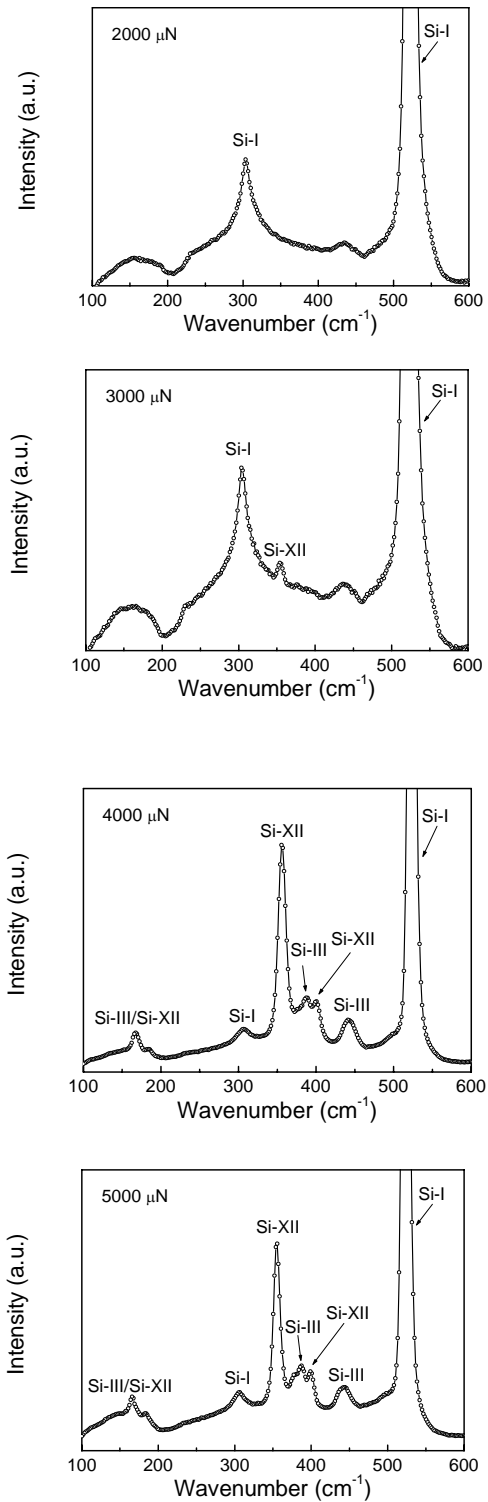

Fig.2 Raman spectra of nanoindentations in c-Si

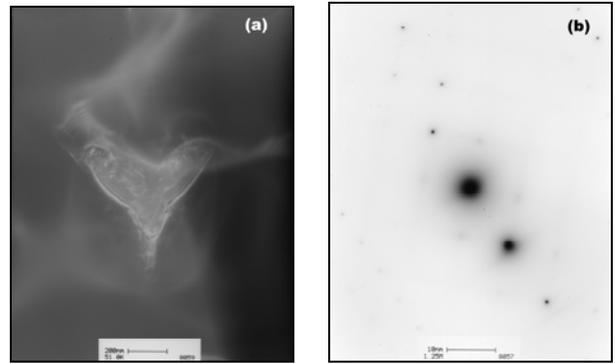

Fig.3 (a) Bright field (BF) TEM images of nanoindentations in c-Si at 3000 μN
(b) The SAD pattern from this indent

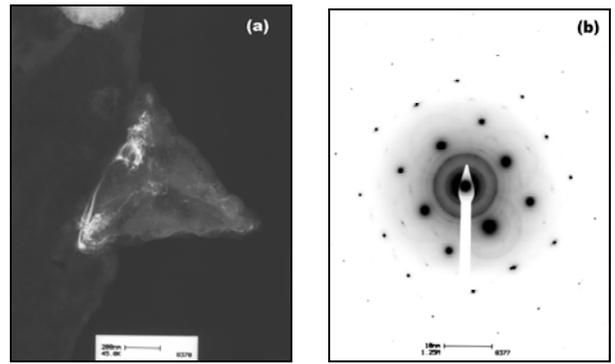

Fig.4 (a) Bright field (BF) TEM images of nanoindentations in c-Si at 5000 μN
(b) The SAD pattern from this indent

## 4. CONCLUSION

In conclusion we have observed nanoindentation-induced phase transformation in c-Si resulting in the end phases Si-III and Si-XII, even though no "pop-in" and "pop-out" are observed in nanoindentation curves. HPP began to appear in c-Si when the maximum load up to 3000 μN. TEM images have been revealed that the optimization HPP transformation in c-Si at the nanoscale occurred when the maximum load applied at 5000 μN. The results indicate that further investigation is needed to understand the nanoindentation behavior of Si.